\nonstopmode

\DeclareRobustCommand{\THEOREMENVS}{
  \newtheorem{theorem}{Theorem}
  
  \newtheorem{lemma}[theorem]{Lemma}
  \newtheorem{corollary}[theorem]{Corollary}

}
\newcommand{\qed}{\ensuremath{\Box}}

\newenvironment{proof*}[1][]{\noindent\ifthenelse{\equal{#1}{}}{\textit{Proof.}}{\textit{Proof #1.}}\hspace{2ex}}{\bigskip}

\newenvironment{caselist}{%
  \begin{list}{\textit{Case}}{}%
}{\end{list}%
}

\newcommand{\nextcase}{\item~}

\newcommand{\ie}{i.\,e.}

\newcommand{\paradot}[1]{\paragraph*{#1.}}

\newcommand{\liftA}[1][]{\tliftA_{#1}}
\newcommand{\tliftA}{\mathsf{liftA}}
\newcommand{\compose}[2][]{\circ^{#1}_{#2}}
\newcommand{\tid}{\mathsf{id}}
\newcommand{\tfmap}{\mathsf{fmap}}
\newcommand{\fmap}[1]{\tfmap\,#1\,}
\newcommand{\tpure}{\mathsf{pure}}
\newcommand{\pure}{\tpure\,}
\newcommand{\law}[1]{\ensuremath{\mbox{\textbf{#1}}}}
\newcommand{\tApplicative}{\ensuremath{\mathsf{Applicative}}}
\newcommand{\Applicative}{\tApplicative\,}
\newcommand{\idapp}{⊛}
\newcommand{\ulaw}[1]{\ensuremath{\mbox{\textit{#1}}}}

\DeclareRobustCommand{\TITLE}{%
  Equivalence of Applicative Functors and Multifunctors
}
\DeclareRobustCommand{\DATE}{February 2022, January 2024
}
\DeclareRobustCommand{\ABSTRACT}{
  McBride and Paterson \citeyearpar{mcBridePaterson:jfp08}
  introduced \texttt{Applicative} functors to
  Haskell, which are equivalent to the lax monoidal functors (with
  strength) of category theory.  Applicative functors $F$ are presented
  via \emph{idiomatic application} $\_{⊛}\_ : F\,(A → B) → FA → FB$ and
  laws that are a bit hard to remember.
  \citet{capriottiKaposi:msfp14} observed that applicative functors
  can be conceived as multifunctors, \ie, by a family $\liftA[n] : (A_1 \to \dots
  \to A_n \to C) \to FA_1 \to \dots \to FA_n \to FC$ of
  \texttt{zipWith}-like functions that generalize \texttt{pure} $(n=0)$,
  \texttt{fmap} $(n=1)$ and \texttt{liftA2} $(n=2)$.  This reduces the
  associated laws to just the first functor law and
  a uniform scheme of second (multi)functor laws, \ie, a composition
  law for $\liftA$.
  In this note, we rigorously prove that applicative functors are
  in fact equivalent to multifunctors, by interderiving their laws.
}
\DeclareRobustCommand{\INTRO}{%

\section{Introduction}

\citet{mcBridePaterson:jfp08} introduce applicative functors as a
Haskell type constructor class $\Applicative F$ with two methods
\[
\begin{array}{lll@{\qquad}l}
  \tpure & : & A \to F A
    & \mbox{embedding} \\
  \_\idapp\_ & : & F\,(A \to B) \to F A \to F B
    & \mbox{idiomatic application, left associative} \\
\end{array}
\]
satisfying four laws:
\[
\begin{array}{l@{\qquad}lll}
  \ulaw{identity}
    & \pure (\lambda x \to x) \idapp u & = & u
\\
  \ulaw{composition}
    & \pure (\lambda f g x \to f\,(g\,x)) \idapp u \idapp v \idapp w
    & = & u \idapp (v \idapp w)
\\
  \ulaw{interchange}
    & \pure (\lambda f \to f x) \idapp u
    & = & u \idapp \pure x
\\
  \ulaw{homomorphism}
    & \pure (f\,x)
    & = & \pure f \idapp \pure x
\\
\end{array}
\]
Using the usual definitions of identity $\tid$ and composition $(\_{\circ}\_)$
the first two laws can be presented as follows:
\[
\begin{array}{l@{\qquad}lll}
  \ulaw{identity}
    & \pure \tid \idapp u & = & u
\\
  \ulaw{composition}
    & \pure (\_{\circ}\_) \idapp u \idapp v \idapp w
    & = & u \idapp (v \idapp w)
\end{array}
\]

Functoriality of $F$ is recovered via $\fmap f u = \pure f \idapp u$
where \ulaw{identity} acts as the first functor law.  The second
functor law can be derived via \ulaw{composition} and
\ulaw{homomorphism} as follows:
\[
\begin{array}{llllll}
  \fmap f (\fmap g u)
    & = & \pure f \idapp (\pure g \idapp u)
    & = & \pure (\_{\circ}\_) \idapp \pure f \idapp \pure g \idapp u \\
    & = & \pure (f \circ\_) \idapp \pure g \idapp u
    & = & \pure (f \circ g) \idapp u \\
    & = & \fmap {(f \circ g)} u
\end{array}
\]

Unfortunately, \citeauthor{mcBridePaterson:jfp08}'s laws are not easy
to remember, especially the \ulaw{composition} and \ulaw{interchange}
laws.  They do not follow simple patterns like the functor laws which
can be seen as actions of the function category, or the monad laws,
which can be conceived as generalization of the monoid laws.  It is
also not intuitively clear at a glance that these laws are complete.

Starting with GHC 8.2 (2017), $\tApplicative$s can also be given via
$\liftA[2] : (A \to B \to C) \to FA \to FB \to FC$
rather than idiomatic application, which are
interdefinable:
\[
\begin{array}{lll}
  \liftA[2]\,f\,u\,v & = & \pure f \idapp u \idapp v \\
  h \idapp u         & = & \liftA[2]\,(\lambda f x \to f x)\,h\,u \\
\end{array}
\]
However, to this date (2024-01-24) the documentation\footnote{
  \url{https://hackage.haskell.org/package/base-4.19.0.0/docs/Control-Applicative.html}
}
of $\tApplicative$ does not spell out the type class laws in terms of
$\liftA[2]$.

Note that $\liftA[2]$ appears to be the \emph{binary} generalization
of the \emph{unary} $\tfmap = \liftA[1] : (A \to B) \to FA \to FB$.
In the same way, we get the \emph{nullary} $\tpure = \liftA[0] : A \to
FA$.

In this note, we show that the further generalization to arbitrary
arities $\liftA[n]$ gives very elegant laws for the family
$\liftA[n]$, which are just generalizations of the two functor laws.

The infinite family $\liftA[n]$ can be truncated to $n \leq 2$,
yielding the following composition laws in addition to the functor
laws (for $\liftA[1]$):
\[
\begin{array}{lll@{\quad}l}
  \liftA[1]\,f\,(\liftA[0]\,x) & = & \liftA[0]\,(f\,x) & \mbox{\ulaw{homomorphism}} \\
  \liftA[2]\,f\,(\liftA[0]\,x) & = & \liftA[1]\,(f\,x) & \mbox{\ulaw{homomorphism}} \\
  \liftA[2]\,f\,u\,(\liftA[0]\,y) & = & \liftA[1]\,(\lambda x \to f\,x\,y)\,u
    & \mbox{\ulaw{exchange}} \\
  \liftA[2]\,f\,(\liftA[1]\,g\,u) & = & \liftA[2]\,(f \circ g)\,u
    & \mbox{2nd functor law} \\
  \liftA[2]\,f\,u\,(\liftA[1]\,h\,v) & = & \liftA[2]\,(\lambda x \to f\,x \circ h)\,u\,v \\
  \liftA[1]\,f\,(\liftA[2]\,g\,u\,v) & = & \liftA[2]\,(\lambda x \to f \circ g\,x)\,u\,v \\
  \liftA[2]\,f\,(\liftA[2]\,g\,u\,v)\,w & = &
    \multicolumn 2 l {
    \liftA[2]\,(\lambda x (y,z) \to f\,(g\,x\,y)\,z)\,u\,(\liftA[2]\,(\_{,}\_)\,v\,w)
    } \\
  \liftA[2]\,f\,u\,(\liftA[2]\,g\,v\,w) & = &
    \liftA[2]\,(\lambda (x,y) \to f\,x \circ g\,y)\,(\liftA[2]\,(\_{,}\_)\,u\,v)\,w
    \\
\end{array}
\]

\section{Applicative Functors as Multifunctors}

\paradot{Preliminaries: generalized composition}
If $f : A_{1..n} \to C \to D$ and $g : B_{1..m} \to C$ then let
$f \compose[m]n g : A_{1..n} \to B_{1..m} \to D$ be
defined by
\[
  (f \compose[m]n g)\;a_{1..n}\;b_{1..m} = f\,a_{1..n}\,(g\,b_{1..m})
  .
\]
Herein, $h\,x_{1..n}$ is to be understood as curried application
$h\,x_1\,\dots\,x_n$.  A sequence $x_{1..n}$ may more succinctly be
written as $\vec x^n$ or just $\vec x$.

Note that $(f \compose[1]0 g)\,x = f\,(g\,x)$ is ordinary unary function composition.
Further, $(f \compose[0]n y)\,a_{1..n} = f\,a_{1..n}\,y$ is partial application of $f$ to its $n+1$st argument, which for $n=0$ is just plain application: $f \compose[0]0 y = f\,y$.
\paradot{Multifunctors}
To distinguish our concept of applicative functors from that of
\citeauthor{mcBridePaterson:jfp08}, we temporarily call them
\emph{multifunctor}.\footnote{
  The name \emph{multi-functor} is taken from
  \citet{capriottiKaposi:msfp14} and already motivated there as means
  to ``naturally arrive at the definition of the $\tApplicative$ clase
  via an obvious generalization of the notion of functor.''
}

A multifunctor $F$ shall be witnessed by a family of functions ($n \ge 0$)
\[
  \liftA[n] :
    (A_1 \to \dots \to A_n \to C) \to FA_1 \to \dots \to FA_n \to FC
\]
satisfying the following laws:
\[
  \begin{array}{l@{\qquad}l@{~~}l@{~~}l}
\law{identity} & \liftA[1]\,\tid & = & \tid \\
\law{composition}
  & \liftA[n+1+m]\,f\,\vec u^n\,(\liftA[k]\,g\,\vec v^k)
  & =
  & \liftA[n+k+m]\,(f \compose[k]n g)\,\vec u\, \vec v
  \end{array}
\]
We may % take the liberty of writing a sequence $u_{1..n}$ as just
drop the index to $\tliftA$ when it is
generic or clear from the context of discourse.

Just functoriality of $F$ can be recovered by $\tfmap = \liftA[1]$ with \law{identity} being the first functor law and \law{composition} specializing to the second functor law with $n=m=0$ and $k=1$:
\[
  \liftA[1]\,f\,(\liftA[1]\,g\,v) = \liftA[1]\,(f \compose[1]0 g)\,v
\]
Pure computations are represented via $\tpure = \liftA[0]$, with the composition law specializing to:
\[
  \liftA\,f\,\vec u^n\,(\pure x) %\,\vec w
  =
  \liftA\,(f \compose[0]n x)\,\vec u %\,\vec w
\]
For $\tfmap$ ($n = m = 0$) this yields $\fmap f (\pure x) = \pure (f\,x)$.
For just $n=0$ we get law $\liftA[1+m]\,f\,(\pure x)\,\vec w =
\liftA[m]\,(f\,x)\,\vec w$. This can be iterated to
$\liftA[n]\,f\,(\tpure\,x_1)\,\dots\,(\tpure\,x_n) =
\tpure\,(f\,x_{1..n})$ corresponding to the intuition that composition
of effect-free computations is again an effect-free computation.

\subsection{Multifunctors are applicative}

Idiomatic application can be obtained as a special case of
$\liftA[2]$:
\[
\begin{array}{lll}
  \_\idapp\_ & : & F (A \to B) \to F A \to F B \\
  u \idapp v & = & \liftA[2]\,\tid\,u\,v
\end{array}
\]
We easily derive its laws:
\begin{enumerate}

\item \ulaw{identity}:
\[
\begin{array}{lllllll@{\qquad}l}
  \pure \tid \idapp u
    & = & \liftA[2]\,\tid\,(\pure \tid)\, u
    & = & \liftA[1]\,(\tid \compose[0]{0} \tid)\, u
    &&
    & \mbox{by \law{composition}}
\\
    & = & \liftA[1]\,(\tid \, \tid)\, u
    & = & \liftA[1]\,\tid\, u
    & = & u
    & \mbox{by \law{identity}}
\\
\end{array}
\]

\item \ulaw{composition}.
\[
\begin{array}{lllll@{\qquad}l}
\multicolumn{6}{l}{
  \pure (\_{\circ}\_) \idapp u \idapp v \idapp w
}
\\
    & = & \liftA[2]\,\tid\,(\pure (\_{\circ}\_)\, u \idapp v \idapp w
    & = & \liftA[1]\,(\_{\circ}\_)\,u \idapp v \idapp w
    & \mbox{by \law{composition}}
\\
    & = & \liftA[2]\,\tid\,(\liftA[1]\,(\_{\circ}\_)\,u)\,v \idapp w
    & = & \liftA[2]\,(\_{\circ}\_)\,u\,v \idapp w
    & \mbox{by \law{composition}}
\\
    & = & \liftA[2]\,\tid\,(\liftA[2]\,(\_{\circ}\_)\,u\,v)\,w
    & = & \liftA[3]\,(\_{\circ}\_)\,u\,v\,w
    & \mbox{by \law{composition}}
\\
    & = & \liftA[3]\,(\lambda f g x \to f\,(g x))\,u\,v\,w
    & = & \liftA[3]\,(\lambda f g x \to \tid\,f\,(\tid\,g\,x))\,u\,v\,w
\\
    & = & \liftA[3]\,(\tid \compose[2]1 \tid)\,u\,v\,w
    & = & \liftA[2]\,\tid\,u\,(\liftA[2]\,\tid\,v\,w)
    & \mbox{by \law{composition}}
\\
    & = & u \idapp (v \idapp w)
\end{array}
\]

\item \ulaw{homomorphism}.  This has been shown before, here again step-by-step:
\[
\begin{array}{lllll@{\qquad}l}
  \pure f \idapp \pure x
    & = & \liftA[2]\,\tid\,(\pure f)\,(\pure x)
    & = & \liftA[1]\,(\tid\,f)\,(\pure x)
    & \mbox{by \law{composition}}
\\
    & = & \liftA[1]\,f\,(\pure x)
    & = & \liftA[0]\,(f\,x)
    & \mbox{by \law{composition}}
\\
    & = & \pure\,(f\,x)
\end{array}
\]

\item \ulaw{interchange}:
\[
\begin{array}{lllll@{\qquad}l}
  u \idapp (\pure x)
    & = & \liftA[2]\,\tid\,u\,(\liftA[0]\,x)
    & = & \liftA[1]\,(\tid \compose[0]{1} x)\,u
    & \mbox{by \law{composition}}
\\
    & = & \liftA[1]\,(\lambda f \to f\,x)\,u
    & = & \pure (\lambda f \to f x) \idapp u
    & \mbox{by \law{composition}}
\end{array}
\]

\end{enumerate}

\subsection{Applicative functors are multifunctors}

Following \citet{mcBridePaterson:jfp08}, the family $\liftA[n]$ can be
defined for each applicative functor:
\[
\begin{array}{lll}
  \liftA[0]  \,x & = & \pure x \\
  \liftA[n+1]\,f\,\vec u\,v & = & \liftA[n]\,f\,\vec u \idapp v
\end{array}
\]
The \law{identity} is just \ulaw{identity}. We establish
\law{composition} by a series of inductions.

\begin{lemma}[Frame]
  \label{lem:frame}
  If\/ $\liftA[n]\,f\,\vec u^n = \liftA[k]\,g\,\vec v^k$ then
  $\liftA[n+m]\,f\,\vec u\,\vec w^m = \liftA[k+m]\,g\,\vec v\,\vec w$.
\end{lemma}
\begin{proof}
  By induction on $m$.
\end{proof}

As a consequence of \Cref{lem:frame}, we only need to show the composition law for $m=0$:
\[
  \liftA[n+1]\,f\,\vec u^n\,(\liftA[k]\,g\,\vec v^k)
  =
  \liftA[n+k]\,(f \compose[k]n g)\,\vec u\, \vec v
\]
We first show the case $n=0$:
\begin{lemma}[Composition for $n=0$]
\label{lem:compnzero}
\[
  \liftA[1]\,f\,(\liftA[k]\,g\,\vec v) = \liftA[k]\,(f \compose[k]0 g)\,\vec v
\]
\end{lemma}
\begin{proof*}
By induction on $k$.
\begin{caselist}
\nextcase $k=0$: This is \ulaw{homomorphism}.
\nextcase $k \to k+1$.
\[
\begin{array}{lll@{\qquad}l}
\multicolumn 4 l {
  \liftA[1]\,f\,(\liftA[k+1]\,g\,\vec v\,w)
}\\
    & = & \pure f \idapp (\liftA[k]\,g\,\vec v \idapp w)
\\
    & = & \pure (\_{\circ}\_) \idapp \pure f \idapp \liftA[k]\,g\,\vec v \idapp w
    & \mbox{by \ulaw{composition}}
\\
    & = & \pure (f \circ \_) \idapp \liftA[k]\,g\,\vec v \idapp w
    & \mbox{by \ulaw{homomorphism}}
\\
    & = & \liftA[k]\,((f \circ \_) \compose[k]0 g)\,\vec v \idapp w
    & \mbox{by ind.hyp.}
\\
    & = & \liftA[k+1]\,(f \compose[k+1]0 g)\,\vec v\, w
\end{array}
\]
For the last step, note that
\(
  (f \circ \_) \compose[k]0 g
  = \lambda \vec x^k \to (f \circ \_) (g\,\vec x)
  = \lambda \vec x \to f \circ (g\,\vec x)
  = \lambda \vec x y \to f (g\,\vec x\,y)
  = f \compose[k+1]0 g
\).
\qed
\end{caselist}
\end{proof*}

\begin{corollary}[Composition for $k=0$]
\label{cor:compkzero}
\[
  \liftA[n+1]\,f\,\vec u\,(\pure x)
  = \liftA[n]\,(f \compose[0]n x)\,\vec u
\]
\end{corollary}
\begin{proof*}
\[
\begin{array}{lllll@{\qquad}l}
\multicolumn 6 l {
  \liftA[n+1]\,f\,\vec u\,(\pure x)
}\\
    & = & \liftA[n]\,f\,\vec u \idapp \pure x
    & = & \pure (\lambda k \to k\,x) \idapp \liftA[n]\,f\,\vec u
    & \mbox{by \ulaw{exchange}}
\\
    & = & \liftA[n]\,((\lambda k \to k\,x) \compose[n]0 f)\,\vec u
    & = & \liftA[n]\,(f \compose[0]n x)\,\vec u
\end{array}
\]
The last step is justified by
\(
    (\lambda k \to k\,x) \compose[n]0 f
  = \lambda \vec y^k \to (\lambda k \to k\,x)\,(f \vec y)
  = \lambda \vec y^k \to f \vec y\,x
  = f \compose[0]n x
\).
\qed
\end{proof*}

\begin{theorem}[Composition]
  \label{thm:comp}
\[
  \liftA[n+1]\,f\,\vec u^n\,(\liftA[k]\,g\,\vec v^k)
  =
  \liftA[n+k]\,(f \compose[k]n g)\,\vec u\, \vec v
\]
\end{theorem}
\begin{proof}
By induction on $k$.
\begin{caselist}

\nextcase $k=0$: This is \Cref{cor:compkzero}.

\nextcase $k \to k+1$:
\[
\begin{array}{lll@{\qquad}l}
\multicolumn 4 l {
  \liftA[n+1]\,f\,\vec u^n\,(\liftA[k+1]\,g\,\vec v^k\,w)
}\\
  & = & \liftA[n]\,f\,\vec u \idapp (\liftA[k]\,g\,\vec v \idapp w)
\\
  & = & \pure (\_{\circ}\_) \idapp \liftA[n]\,f\,\vec u \idapp \liftA[k]\,g\,\vec v \idapp w
  & \mbox{by \ulaw{composition}}
\\
  & = & \liftA[n]\,((\_{\circ}\_) \compose[n]0 f) \,\vec u \idapp \liftA[k]\,g\,\vec v \idapp w
  & \mbox{by \Cref{lem:compnzero}}
\\
  & = & \liftA[n+k]\,(((\_{\circ}\_) \compose[n]0 f) \compose[k]n g) \,\vec u \,\vec v \idapp w
  & \mbox{by ind.hyp.}
\\
  & = &
   \liftA[n+k+1]\,(f \compose[k+1]n g)\,\vec u\, \vec v\, w
\end{array}
\]
For the last step, we calculate
\(
  ((\_{\circ}\_) \compose[n]0 f) \compose[k]n g
  = \lambda \vec x^n \vec y^k \to (\lambda \vec x^n \to (f \vec x) \circ \_))\,\vec x\, (g\,\vec y)
  = \lambda \vec x^n \vec y^k \to (f \vec x) \circ (g\,\vec y)
  = \lambda \vec x^n \vec y^k z \to f \vec x \, (g\,\vec y\,z)
  = f \compose[k+1]n g
  .
\)
\end{caselist}
\end{proof}

}
\DeclareRobustCommand{\CONCLUSION}{%
  Q.E.D.
\bibliographystyle{abbrvnat}
\bibliography{auto-Applicative}
}

\nonstopmode

\documentclass[a4paper]{scrartcl}

\usepackage{ifthen}
\usepackage{iftex}
\usepackage{natbib}
\usepackage{amsmath}
\usepackage{amsfonts}
\usepackage{amssymb}
\usepackage{amsthm}
\usepackage{cleveref}
\usepackage{textgreek}

\newif\ifxetexorluatex
\ifxetex
  \xetexorluatextrue
\else
  \ifluatex
    \xetexorluatextrue
  \else
    \xetexorluatexfalse
  \fi
\fi

\ifxetexorluatex
  \usepackage{unicode-math}
\fi

\usepackage{newunicodechar}
\newunicodechar{λ}{\ensuremath{\mathnormal\lambda}}
\newunicodechar{ε}{\ensuremath{\mathnormal\varepsilon}}
\newunicodechar{⊛}{\ensuremath{\circledast}}
\newunicodechar{→}{\ensuremath{\to}}

\ifdefined\heading
\else
\newcommand{\heading}[1]{\section{#1}}
\fi

\ifdefined\subheading
\else
\newcommand{\subheading}[1]{\subsection{#1}}
\fi

\ifdefined\THEOREMENVS\THEOREMENVS\fi

\ifdefined\TITLE\title{\TITLE}\fi

\ifdefined\AUTHOR
\author{\AUTHOR}
\else
\author{%
  Andreas Abel
  \\
  Department of Computer Science,
  Gothenburg University,
  Sweden
  }
\fi

\ifdefined\DATE\date{\DATE}\else\date{\today}\fi

\begin{document}

\maketitle

\ifdefined\ABSTRACT
\begin{abstract}
\ABSTRACT
\end{abstract}
\fi

\ifdefined\INTRO\INTRO\fi

\ifdefined\MAIN\input{latex/\MAIN}\fi

\ifdefined\CONCLUSION\CONCLUSION\fi

\end{document}